\documentclass[a4paper,11pt]{article}
\usepackage{pos}
\usepackage{multirow}

\title{Physics perspectives of a CMS near-beam proton spectrometer at the HL-LHC}
\ShortTitle{PPS forward physics upgrade}

\author*[a]{Michael Pitt}
\onbehalf{on behalf of the CMS collaboration}
\affiliation[a]{The University of Kansas, Department of Physics,
Lawrence, USA}

\emailAdd{michael.pitt@cern.ch}

\abstract{The High-Luminosity Large Hadron Collider (HL-LHC) is designed to achieve higher instantaneous luminosities, enabling the exploration of the rarest processes of the Standard Model (SM). The CMS collaboration has published an Expression of Interest to pursue the study of the central exclusive production processes, using near-beam detectors. This report details both the expected performance and the scientific potential of the CMS near-beam proton spectrometer at the HL-LHC.
}

\FullConference{The Eleventh Annual Conference on Large Hadron Collider Physics (LHCP2023)\\
 22-26 May 2023\\
 Belgrade, Serbia\\}

\begin{document}
\maketitle

\section{Introduction}

In hard proton--proton collisions at the LHC, final-state particles are produced predominantly from the interactions of quarks or gluons that originate from the colliding protons. These events typically involve the production of a hard process accompanied by many energetic particles, which arise from proton remnants, initial-state radiation, or multiparton interactions. In rare occasions, particles can be produced by exchanging color-neutral mediators, such as photons in electromagnetic interactions or pomerons in strong interactions, allowing the protons to emerge intact but lose a fraction of their momentum. Hard processes associated with an intact proton cover a broad range of topologies and scales. The coherent photoproduction of vector mesons has so far been measured in ultraperipheral proton--lead collisions without proton tagging targeting scales mainly up to a few GeV~\cite{ALICE:2014eof,ALICE:2018oyo,CMS:2018bbk,CMS:2019awk}. The Mueller--Tang jet topology with two energetic jets with a transverse momentum above tens of GeV with and without proton tagging was measured by the TOTEM and CMS collaborations~\cite{TOTEM:2021rix}. The central exclusive production (CEP) of final states with masses of a few GeV, without proton tagging, was reported by the LHCb collaboration \cite{LHCb:2013nqs,LHCb:2016oce}. At the same time, hard scattering events were studied by ATLAS \cite{ATLAS:2020mve,ATLAS:2022uef} and CMS \cite{CMSCollaborationTOTEM:2023dbs,CMS:2018uvs,CMS:2023roj,CMS:2022zfd,CMS:2022dmc,CMS:2023naq} utilizing proton tagging techniques. The CEP processes at masses of the order of a few hundred  GeV are predominantly produced by fusion of two photons. This implies that the LHC can be treated as a photon collider, offering a new means for probing fundamental physics and searching for new particles.

Tagging events with two intact protons can discriminate between inclusive and exclusive processes produced in proton--proton collisions. These protons, having lost a fraction of their initial momentum (commonly denoted by $\xi=\Delta p/p$), will be laterally deflected from the proton beam by the LHC magnets and can be measured using dedicated near-beam detectors. In typical LHC runs, several tens of proton--proton interactions occur per proton bunch crossing. In a small percentage of these interactions, at least one proton undergoes diffractive scattering with a large momentum transfer and remains intact. Consequently, an inclusive process accompanied by a pair of protons resulting from such additional interactions (pileup) can resemble the exclusive final state. Hence, CEP studies face the significant challenge of a large combinatorial background (i.e., an inclusively produced central system with pileup protons detected on both sides). 

The Time-of-Flight (ToF) method can be used to measure the arrival times of protons, allowing the reconstruction of the spatial or temporal coordinates of the vertex associated with the scattered protons \cite{Cerny:2020rvp}. By aligning the kinematic properties of the protons with those of the central system and central production vertex, it becomes feasible to identify CEP events at the LHC.

\section{The CMS Precision Proton Spectrometer }

CEP events with intact protons can be tagged using specialized near-beam proton spectrometers. These detectors are housed in a movable detector vessel, enabling them to approach the beam center within just a few millimeters and measure the scattered protons. Introduced in 2016 as a joint effort between CMS and TOTEM collaborations \cite{CMS:2014sdw}, the CMS Precision Proton Spectrometer (PPS) has since been successfully operational as a standard CMS subdetector \cite{CMS:2023gfb}. The PPS detector, located about 200 meters from the CMS interaction point in both arms, has tracking and timing stations that allow for precise measurement of forward protons. 

Prior to the transition to the High-Luminocity LHC (HL-LHC) phase, the accelerator layout will be rearranged, and the CMS collaboration proposes a new detector design for the PPS \cite{CMS:2021ncv}. The new locations of the forward detectors are selected at $z=\pm196$, $\pm220$, and $\pm234$ meters from the CMS interaction point, making them sensitive to protons with a momentum loss from 1.5\% to 20\%, which corresponds to a mass range of centrally produced final-state particles between 133 GeV and 2.7 TeV. Additionally, with a further station positioned at $z=\pm420$ meters, the $\xi$ acceptance range will be as low as 0.325\%, including central masses as light as 43 GeV.

\section{Physics with the PPS sub-detector at the HL-LHC}

\subsection{Standard Model processes}

The HL-LHC upgrade aims to deliver about 3000 fb$^{-1}$ of total integrated luminosity over a span of 10-12 years, enabling the study of the rarest SM processes, including CEP. Table \ref{tab:xsec-sm-processes} shows the fiducial cross-section for different SM photon-induced CEP processes in $pp$ collisions at $\sqrt{s}=14$~TeV, where one of the two protons will be within the PPS acceptance for two scenarios: without the 420m station ($\xi\in\left(0.0142,0.1967\right)$), and with all four stations ($\xi\in\left(0.00325,0.1967\right)$).

\begin{table}[h!]
  \begin{center}
    \caption{Fiducial cross-sections of exclusive photon-initiated production processes in pp collisions at $\sqrt{s}=14$\,TeV with one or both protons within the PPS acceptance for two scenarios, with and without the station at $\pm420$~m.}

    \label{tab:xsec-sm-processes}
    \begin{tabular}{|l|c|c|c|c|}
    \hline
    \multirow{3}{*}{\textbf{Process}} & \multicolumn{4}{c|}{\textbf{fiducial cross section [fb]}} \\ 
            
      & \multicolumn{2}{c}{\textbf{2 tag}} & \multicolumn{2}{|c|}{\textbf{1 tag}} \\ \hline

      & \textbf{w/o 420} & \textbf{all stations}  & \textbf{w/o 420} & \textbf{all stations} \\   
        
      \hline

$\rm jj$ & 2 &  60 & 219  & 526 \\
$\rm b\bar{b}$ &  0.04 &  1.7 & 6.3 &  15 \\
$W^+W^-$ & 15 &  37  & 152 & 178 \\
$\mu\mu$ & 1.3 &  46 & 172 &  417  \\
$\rm t\bar{t}$ & 0.1 &  0.15 & 0.65 &  0.74 \\
H & 0 &  0.07 & 0.23 &  0.30 \\
$HW^+W^-$ & 0.01 &  0.01 & 0.06 & 0.07 \\
$ZZ$ & 0.03 &  0.06 & 0.23 &  0.26 \\
$Z\gamma$ & 0.02 &  0.04 & 0.15 &  0.17 \\
$\gamma\gamma$ & 0.003 &  0.02 & 0.19 &  0.33 \\

    \hline
    \end{tabular}
  \end{center}
\end{table}

The cross-section was calculated with the MadGraph5\_aMC@NLO event generator~\cite{Alwall:2014hca}, using the Equivalent Photon Approximation (EPA) for elastic photon fluxes~\cite{Budnev:1974de}, assuming a gap survival probability of 90\% and using the  \textit{MMHT2015qed\_nlo\_inelastic} PDF set for inelastic photon fluxes~\cite{Harland-Lang:2019pla}, assuming a gap survival probability of 70\%. The production of the Single Higgs boson was computed in the Higgs Effective Field Theory (HEFT) model~\cite{Shifman:1979eb,Dawson:1993qf,Kniehl:1995tn}, and the loop-induced production of neutral di-boson final states ($ZZ$, $Z\gamma$, and $\gamma\gamma$) was generated using the \textit{loop\_qcd\_qed\_sm} model~\cite{Hirschi:2015iia}. A central detector selection cut of $p_T>20$\, GeV, and $|\eta|<2.5$ on the generated objects was applied for processes with two particles in the final state.

Despite the relatively small cross-sections, the amount of data delivered by the HL-LHC will enable the exploration of various SM CEP processes. This includes the exclusive production of QCD multijet events~\cite{Harland-Lang:2015cta}; the di-lepton production for detector calibration studies~\cite{CMS:2022hly} and model tunes; $WW$ production, which is a particularly clean channel in its fully leptonic decay mode~\cite{CMS:2014sdw}; and the exclusive production of top-quark pairs~\cite{Baldenegro:2022kaa,Martins:2022dfg,Goldouzian:2016mrt,Pitt:2023mtm}. The central exclusive Higgs boson production was extensively studied, with a predicted cross-section of the order of a few fb, predominantly in the QCD-initiated production mode~\cite{Khoze:2000cy,Petrov:2003yt,Khoze:2002py,DeRoeck:2002hk,Cudell:2010cj,Maciula:2010tv,Coughlin:2009tr,Ryutin:2012np}. The photon-induced cross-section for this production mode is an order of magnitude lower. For the Higgs boson masses, tagging two protons is feasible only with the 420\,m station scenario. However, the associated production of a Higgs boson and a $\rm W^{+}W^{-}$ vector--boson pair holds the potential for probing the Higgs sector in CEP events, even without the $\pm$420\,m stations, primarily in processes dominated by photon-induced production.

\subsection{Physics Beyond the Standard model}

The PPS offers exceptional sensitivity to anomalous couplings and can probe di-boson events with high invariant masses. It will play a crucial role in the search for  Axion-Like Particles (ALPs) through exclusive di-photon events at high invariant masses~\cite{Baldenegro:2018hng}, particularly in single tagged events, bolstered by recent advancements in the theoretical framework~\cite{Harland-Lang:2022jwn}. Proton tagging is instrumental in investigating the BSM $\gamma\gamma\gamma Z$ coupling in exclusive $Z\gamma$ final states~\cite{Azzi:2019yne}, as well as  anomalous $\gamma\gamma WW$ coupling in exclusive $WW$ events.

The PPS is particularly attractive for searches in supersymmetric scenarios with compressed spectra. An example of this is the process $pp\to\tilde{\ell}\tilde{\ell}\to\ell\ell\tilde{\chi}_1^0\tilde{\chi}_1^0$, where both neutralinos ($\tilde{\chi}_1^0$) are produced at rest. At the LHC, such searches typically require high initial state radiation activity to boost neutralinos and tag events with high missing transverse momentums. In the exclusive production, where both protons are tagged, the di-slepton mass ($m_{\tilde{\ell}\tilde{\ell}}$) is measured by the PPS independently of the outgoing particles reconstructed by the central detector~\cite{Harland-Lang:2018hmi,Beresford:2018pbt}.

\section{Summary}

The PPS subsystem extends the scope of the CMS physics program at HL-LHC in the field of central exclusive production processes, both in terms of the mass range and the higher amount of statistics. With its enhanced timing capabilities, the newly designed PPS meets the challenges posed by the high radiation rates anticipated at the HL-LHC. The CMS collaboration has proposed a staged installation strategy, beginning with installing three stations at about 200 meters from the interaction point at the onset of LHC Run 4 (the HL-LHC phase). The construction of the station at 420 meters from the interaction point is forseen for LHC Run 5 or later, aligning with the evolving requirements and objectives of the experiment.

\end{document}